\documentclass[a4paper,11pt]{article}
\usepackage{pos}

\let\OLDthebibliography\thebibliography
\renewcommand\thebibliography[1]{
  \OLDthebibliography{#1}
  \setlength{\parskip}{0pt}
  \setlength{\itemsep}{-3pt}
\footnotesize
}

\title{
Interglueball potential in lattice SU(N) gauge theories
}

\author*[a,b,c]{Nodoka Yamanaka}
\author[d,c]{Atsushi Nakamura}
\author[e,d]{Masayuki Wakayama}

\affiliation[a]{Department of Physics, Kennesaw State University,\\
Kennesaw, Georgia 30144, USA}

\affiliation[b]{Kobayashi-Maskawa Institute for the Origin of Particles and the Universe, Nagoya University,\\
Furo-cho Chikusa-ku, Nagoya, 464-8602 Japan}

\affiliation[c]{Nishina Center for Accelerator-Based Science, RIKEN,\\
Wako 351-0198, Japan}

\affiliation[d]{Research Center for Nuclear Physics (RCNP), Osaka University,\\
Ibaraki, Osaka 567-0047, Japan}

\affiliation[e]{Department of General Education, Chiba Institute of Technology,\\
Chiba 275-0023, Japan}

\emailAdd{nyamanaka@kmi.nagoya-u.ac.jp}

\abstract{
The dynamics of the glueballs is important in the context of the experimental search as well as for understanding the non-Abelian gauge theory.
The glueballs of the dark $SU(N_c)$ Yang-Mills theory are also good candidates of dark matter.
In this proceedings contribution, we report on the result of the lattice calculation of the interglueball potential of the Yang-Mills theory with the color numbers $N_c=2,3,4$, with a detailed inspection of the systematics due to the discretization.
}

\FullConference{%
 The 38th International Symposium on Lattice Field Theory, LATTICE2021
  26th-30th July, 2021
  Zoom/Gather@Massachusetts Institute of Technology
}

\begin{document}
\maketitle

\section{Introduction}

The $0^{++}$ glueball is the lightest particle of the spectrum of the Yang-Mills theory (YMT) \cite{Mathieu:2008me,Llanes-Estrada:2021evz}.
In quantum chromodynamics (QCD), its existence is still not clear due to the mixing with quarks \cite{Cheng:2006hu}, and the experimental search is currently actively performed, with several candidates such as the $f_0(1710)$ meson \cite{BESIII:2013qqz,Belle:2013eck,Llanes-Estrada:2021evz}.
The glueball of $SU(N_c)$ YMT is also a good candidate of dark matter (DM) \cite{Bertone:2004pz,Boddy:2014yra,Soni:2016gzf,Kribs:2016cew,Battaglieri:2017aum} which consistently explains many phenomena such as the galactic rotation curve.

In the context of the DM, the low energy scattering cross section among DM particles affects the structure of the scale smaller than the galaxy \cite{Spergel:1999mh,Hertzberg:2020xdn}.
It is therefore useful to calculate this physical quantity in a given model if one wants to explain the DM.
The YMT is difficult to handle with perturbative methods, and lattice calculations are mandatory to extract low energy observables \cite{deForcrand:1984eeq,Teper:1987wt,Albanese:1987ds,Teper:1998kw,Morningstar:1999rf,Bali:2000vr,Ishii:2001zq,Ishii:2002ww,Lucini:2004my,Chen:2005mg,Lucini:2010nv,Gregory:2012hu,Yamanaka:2019aeq,Yamanaka:2019yek,Bennett:2020qtj,Dudal:2021gif}.
It is now possible to calculate the scattering between hadrons on lattice.
In this proceedings contribution, we update our calculation of the interglueball potentials in $SU(2)$, $SU(3)$, and $SU(4)$ YMTs \cite{Yamanaka:2019gak} using the HAL QCD method \cite{Ishii:2006ec,Aoki:2009ji,HALQCD:2012aa} by subtracting the centrifugal force to remove the lattice artifact due to higher partial waves.

\section{Setup of lattice calculation and formalism}

In this work, we simulate $SU(2)$, $SU(3)$, and $SU(4)$ YMTs on lattice with several lattice spacings, where we chose the standard plaquette action.
We generate configurations with the pseudo-heat-bath method.
The lattice spacing is expressed in the unit of the scale parameter $\Lambda$, which is left as a free parameter.
To determine the relation between them, we use the result of the lattice calculation of the string tension \cite{Allton:2008ty,Teper:2009uf}
\begin{eqnarray}
\frac{\Lambda}{\sqrt{\sigma}}
=
0.503(2)(40)+ \frac{0.33(3)(3)}{N_c^2}
.
\label{eq:scalestringtension}
\end{eqnarray}
In Table \ref{table:lattice_spacing}, we show the lattice spacings which are used in this work.

\begin{table}[h]
\begin{center}
\begin{tabular}{cccc}
\hline \hline
$N_c$ & $\beta$ & $a \sqrt{\sigma}$ & $a$ $(\Lambda^{-1})$\\
\hline
2 & 2.1 & 0.608 (16) \cite{Teper:1998kw} & 0.356 (27) \\
 & 2.2 & 0.467 (10) \cite{Teper:1998kw} & 0.273 (20) \\
 & 2.3 & 0.3687 (22) \cite{Teper:1998kw} & 0.216 (15) \\
 & 2.4 & 0.2660 (21) \cite{Teper:1998kw} & 0.156 (11) \\
 & 2.5 & 0.1881 (28) \cite{Teper:1998kw} & 0.110 (8) \\
\hline
3 & 5.5 & 0.5830 (130) \cite{Teper:1998kw} & 0.315 (24) \\
 & 5.7 & 0.3879 (39) \cite{Teper:1998kw} & 0.209 (16) \\
 & 5.9 & 0.2613 (28) \cite{Teper:1998kw} & 0.141 (11) \\
 & 6.1 & 0.1876 (12) \cite{Teper:1998kw} & 0.101 (8) \\
\hline
4 & 10.789 & 0.2706 (8) \cite{Lucini:2004my} & 0.142 (3) \\
\hline
\end{tabular}
\end{center}
\caption{
Relations between the lattice spacing $a$ and the scale parameter of the YMT $\Lambda$ used in this work.
The combined statistical and systematic errors are also given.
}
\label{table:lattice_spacing}
\end{table}

The $0^{++}$ glueball are defined on lattice by
\begin{equation}
\hspace{-0.15em} 
\phi (t, \vec{x}) 
=
{\rm Re} [
P_{12} (t, \vec{x}) 
+P_{12} (t, \vec{x}+a\vec{e}_3) 
+P_{23} (t, \vec{x}) 
+P_{23} (t, \vec{x}+a\vec{e}_1) 
+P_{31} (t, \vec{x}) 
+P_{31} (t, \vec{x}+a\vec{e}_2) 
]
,
\label{eq:glueballop}
\end{equation}
where $P_{ij}$ ($i,j = 1,2,3$) are the purely spatial plaquette operator belonging to the $i-j$ plane.
We note that we have to subtract the vacuum expectation value of $P_{ij}$ in the actual calculation.

To improve the signal of the glueball on lattice, we use the APE smearing \cite{Albanese:1987ds,Ishii:2001zq,Ishii:2002ww} by maximizing the following quantity
\begin{equation}
{\rm Re\, Tr} [U_i^{(n+1)}(t, \vec{x}) V_i^{(n)\dagger}(t, \vec{x}) ]
, 
\label{eq:smearing}
\end{equation}
obtained after $n$ iterations, with $U_i^{(n)}$ the link variable, and 
\begin{eqnarray}
V_i^{(n)}(t, \vec{x})
&\equiv &
\alpha U_i^{(n)}(t, \vec{x})
+\sum_{\pm j \neq i} U_j^{(n)}(t, \vec{x})
U_i^{(n)}(t, \vec{x}+\vec{e}_j) U_j^{(n)\dagger}(t, \vec{x}+\vec{e}_i)
.
\end{eqnarray}
The real parameter $\alpha $ and $n$ are chosen so as to minimize the effective mass of the two-point correlator.

\begin{figure}[htbp]
\centering
\includegraphics[width=0.5\textwidth,clip]{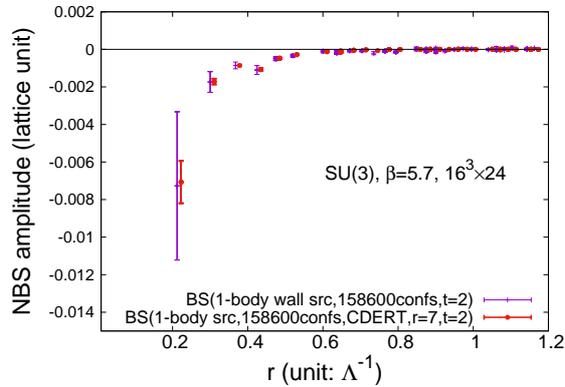}
\caption{\label{fig:su3_beta5p7_glueball_BS_1-body}
NBS amplitude calculated with 1-body sources in $SU(3)$ YMT.
The data were taken at the imaginary time slice $t=2$.
}
\end{figure}

To extract the scattering between glueballs, we define the following Nambu-Bethe-Salpeter (NBS) amplitude:
\begin{equation}
\Psi_{\phi \phi} 
(t,\vec{x}-\vec{y})
\equiv
\frac{1}{V} \sum_{\vec{r}} 
\langle 0 | T[\tilde \phi (t, \vec{x}+\vec{r})\tilde \phi (t, \vec{y}+\vec{r}) {\cal J}(0)] | 0 \rangle
.
\label{eq:NBS}
\end{equation}
Here ${\cal J}$ is the source operator, and it may be chosen arbitrarily as long as it has an overlap with the two-glueball state.
In this work, we choose as ${\cal J}$ the one-glueball operator optimized with the APE smearing as seen above.
We also improve the statistical accuracy of the NBS amplitude by using the cluster decomposition error reduction technique (CDERT) \cite{Liu:2017man}, which consists of restricting the integration in a space where the space-time distances between two glueball operators do not exceed a chosen cutoff so as to only reduce the statistical error and without affecting the physical signal. 
We plot in Fig. \ref{fig:su3_beta5p7_glueball_BS_1-body} the NBS amplitude calculated in the $SU(3)$ YMT.
We may confirm that the use of CDERT greatly improves the signal of the NBS amplitude.

From the NBS amplitude (\ref{eq:NBS}), it is possible to calculate the interglueball potential.
The s-wave interglueball NBS amplitude actually obeys the Schr\"{o}dinger equation in the time-dependent formalism \cite{Ishii:2006ec,Aoki:2009ji,HALQCD:2012aa}
\begin{equation}
\Biggl[
\frac{1}{4m_\phi} \frac{\partial^2}{\partial t^2}-\frac{\partial}{\partial t} + \frac{1}{m_\phi} \nabla^2
-\frac{({\vec r} \times {\vec \nabla})^2}{m_\phi r^2}
\Biggr]
R(t,{\vec r})
=
\int d^3r' U({\vec r},{\vec r}')R(t,{\vec r}')
,
\label{eq:time-dependent_HAL}
\end{equation}
where $R(t,{\vec r}) \equiv \Psi_{\phi \phi} (t,{\vec r}) / e^{-2m_\phi t}$, and $m_\phi$ is the glueball mass fitted from the effective mass of the two-point correlator.
In the low energy limit, the local approximation $U({\vec r},{\vec r}') \approx V_{\phi \phi} (\vec{r}) \delta (\vec{r}-\vec{r}')$ works well.
The crucial advantage of this time-dependent formalism is that the obtained potential will not be affected by excited states so that the computationally costly ground state saturation is not necessary.
We also subtracted the centrifugal force to remove the effect of higher partial waves due to the lattice artifact.

\section{Result}

\begin{figure}[hbt]
\begin{center}
\includegraphics[width=.49\columnwidth]{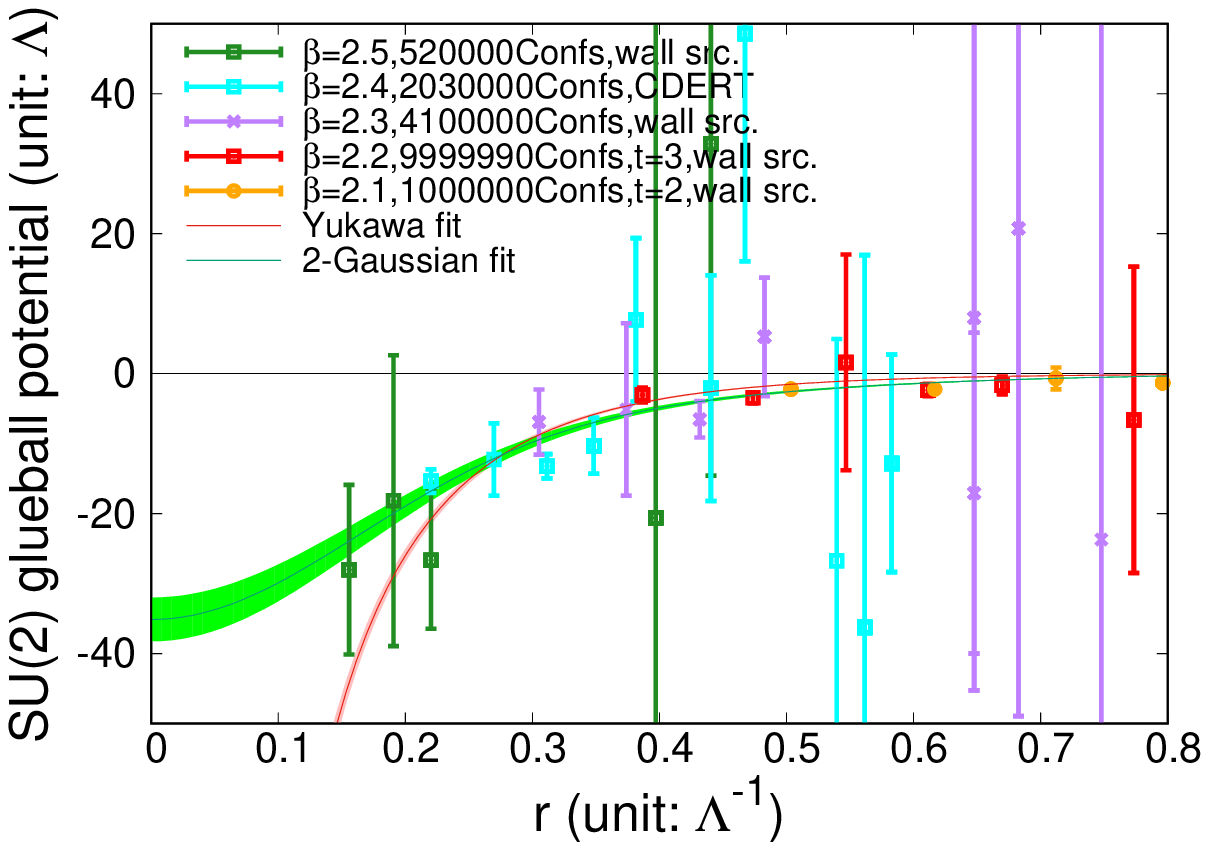}
\includegraphics[width=.49\columnwidth]{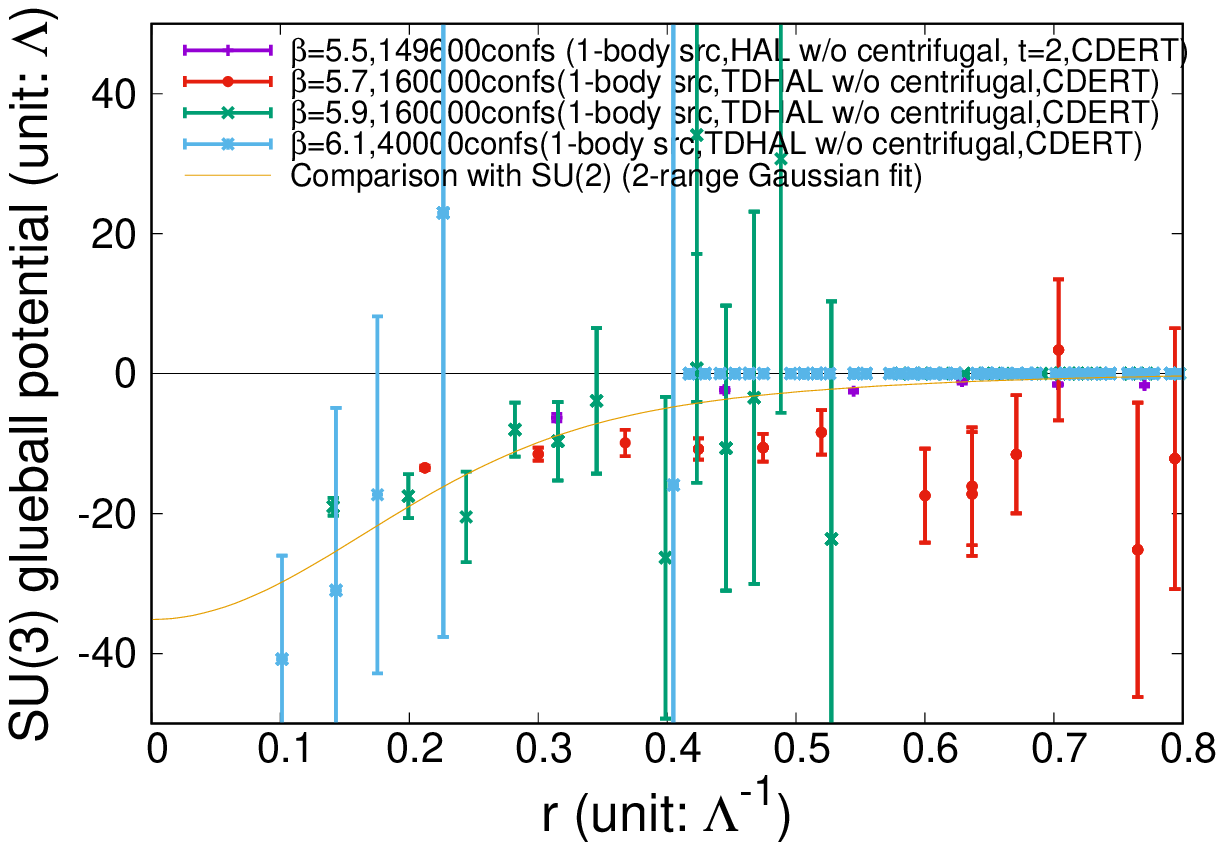}
\includegraphics[width=.49\columnwidth]{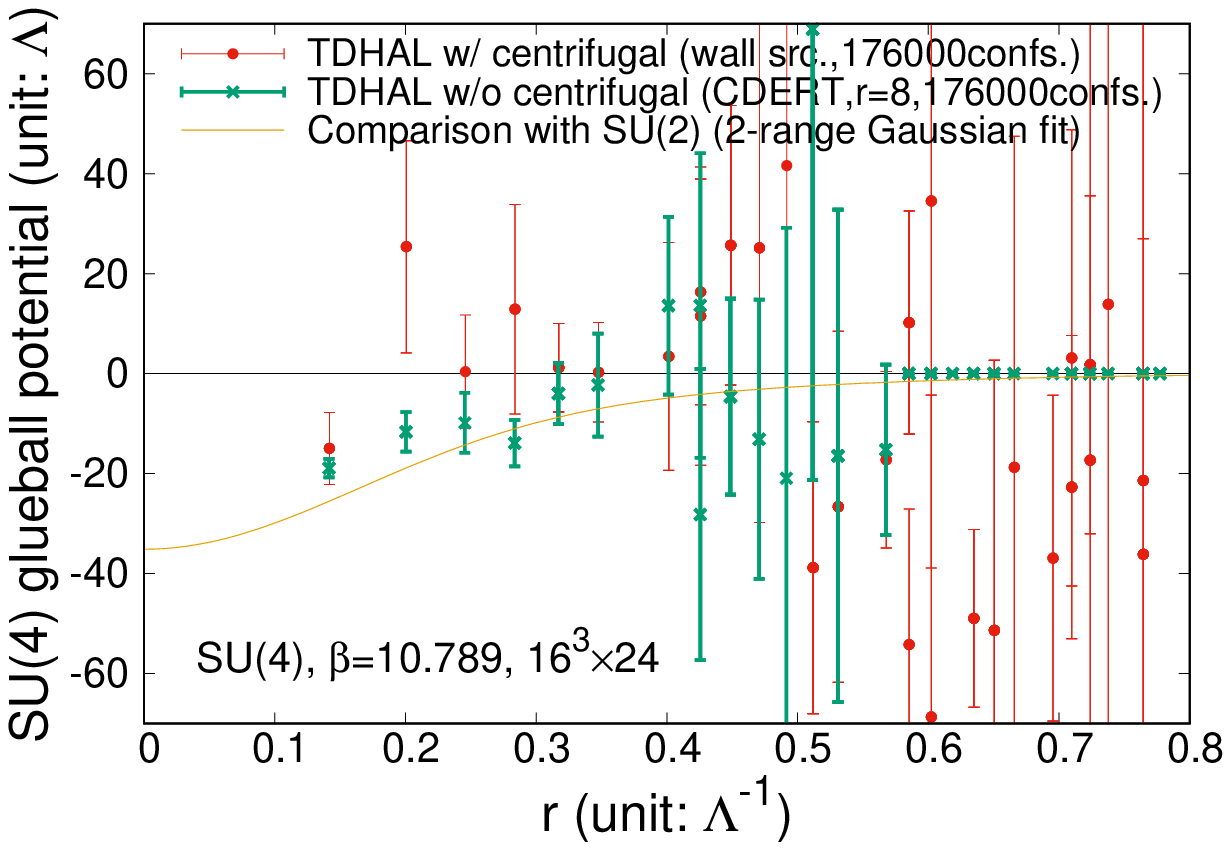}
\caption{
Interglueball potential calculated in the $SU(2,3,4)$ YMTs.
The $SU(2)$ result is fitted with two fitting forms (Yukawa and 2-Gaussian).
For the $SU(3)$ and $SU(4)$ plots, the result of the $SU(2)$ YMT (2-Gaussian fit) is also plotted for comparison.
In the plot of $SU(4)$, we compare the interglueball potential with and without the centrifugal force.
}
\label{fig:glueball_potential}
\end{center}
\end{figure}

Let us first present the result of the $SU(2)$ YMT.
We superpose the results of the calculations with $\beta = 2.1, 2.2, 2.3, 2.4$ and 2.5, and fit the interglueball potential with two fitting forms (Yukawa and 2-Gaussian).
With the Yukawa function, we obtain $V_Y(r) = V_1 \frac{e^{-m_\phi r}}{r}$ where $V_1 = -231\pm 8$ ($\chi^2/$d.o.f. = 1.3).
This result is consistent with the estimation of the strong coupling expansion \cite{Munster:1984zf}.
In the 2-Gaussian fit $V_{2G}(r) = V_1 \frac{e^{-m_\phi r}}{r}+V_2 e^{-\frac{(m_\phi r)^2}{2}}$, we obtain $V_1 = (-8.5 \pm 0.5 ) \Lambda$ and $V_2 = (-26.6 \pm 2.6) \Lambda$ ($\chi^2/$d.o.f. = 0.9).
The result is plotted in the top-left of Fig. \ref{fig:glueball_potential}.

Now we derive the interglueball cross section.
The scattering phase shift may be calculated by solving the following equation
\begin{eqnarray}
\biggl[
\frac{\partial^2}{\partial r^2} 
+k^2
-m_\phi V(r)
\biggr]
\psi (r)
=0
,
\end{eqnarray}
by substituting $V(r)$ with the fitted potential.
The scattering phase shift $\delta$ is given by the asymptotic form of the scattered wave $\psi(r) \propto \sin [kr+\delta(k)]$.
The low energy scattering cross section is then obtained as 
\begin{equation}
\sigma_{\phi \phi} =\lim_{k\to 0}\frac{4 \pi}{k^2} \sin^2 [\delta(k)]
=
\left\{
\begin{array}{ll}
(2.5 - 4.7) \Lambda^{-2} & ({\rm Yukawa \ fit})\cr
(14 - 51) \Lambda^{-2} & ({\rm 2-Gaussian\ fit})\cr
\end{array}
\right.
,
\label{eq:gbcrosssection}
\end{equation}
where the variation corresponds to the statistical error.
The difference between the two results may be interpreted as the systematic error due to the choice of the fitting forms.
From the observation of the shape of the galactic halo \cite{Rocha:2012jg} and galactic collisions \cite{Randall:2007ph}, the scattering cross section of the DM is constrained as $\sigma_{\rm DM} / m_{\rm DM} < 1 \, {\rm cm }^2 /$g.
By equating it with our result (\ref{eq:gbcrosssection}), we obtain the constraint on the scale parameter of the $SU(2)$ YMT $\Lambda > 60 \,{\rm MeV}$.

\begin{figure}[hbt]
\begin{center}
\includegraphics[width=.5\columnwidth]{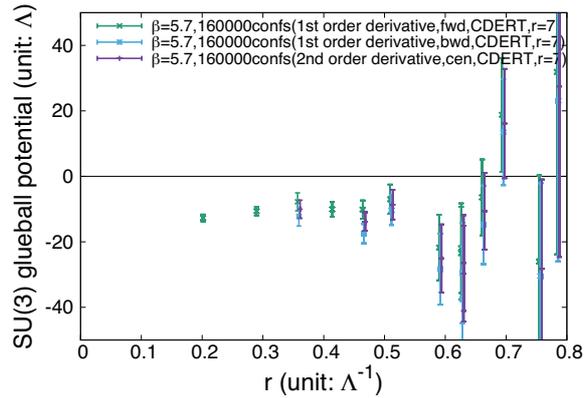}
\caption{
Comparison of the form of finite difference in the calculation of the interglueball potential in $SU(3)$ YMT.
}
\label{fig:glueball_potential_finite_difference}
\end{center}
\end{figure}

Next, we show the result of the calculation of the interglueball potential in the $SU(3)$ YMT.
By superposing the results of $\beta =5.5, 5.7 , 5.9$ and 6.1, we obtain the top-right plot of Fig. \ref{fig:glueball_potential}.
We see that the size of the potential resembles the $SU(2)$ case.
We also see an anomalous flat structure of the potential around $r= 0.4 \Lambda^{-1}$ for $\beta =5.7$.
To inspect the systematics due to the discretization, we changed the form of the finite difference.
We tested with the first order forward, first order backward, and second order central forms, but no significant difference could be seen (see Fig. \ref{fig:glueball_potential_finite_difference}).
However, we see that the difference between the central values is of the same order of magnitude as the values themselves, so the discretization error may be large.
To quantify this point, we have to simulate the YMT with the improved action.

Finally, we also plot the result of the $SU(4)$ YMT with one lattice spacing $\beta = 10.789$ (see the bottom plot of Fig. \ref{fig:glueball_potential_finite_difference}).
This result is also close to the $SU(2)$ one, and the large $N_c$ behavior \cite{Lucini:2001ej,Lucini:2012gg,Hernandez:2020tbc} of the interglueball potential, which should scale as $N_c^{-2}$, is not clear.
In the $SU(4)$ plot of Fig. \ref{fig:glueball_potential}, we also compared the potential with and without the centrifugal force.
We may clearly see the improvement due to the subtraction of the effect of unphysical higher partial waves.

\section{Summary}

In this proceedings contribution, we reported on the recent update of our calculation of the interglueball potential in the $SU(2,3,4)$ YMTs using the so-called HAL QCD method.
We calculated the interglueball potential in these YMTs with several lattice spacings and the scale parameter left as a free variable.
An important improvement we applied in this work is the subtraction of the centrifugal force, which should have no physical effect as long as we are interested in the low energy s-wave scattering, but this is expected to remove the unphysical artifact due to the lattice discretization. 
For the case of $SU(2)$ theory, we could obtain an attractive interglueball potential, and the calculation of the scattering phase shift gave us $\sigma_{\phi \phi} = (2 - 51) \Lambda^{-2}$.
Combining with the constraint on the DM scattering cross section known from observation, we obtain the lower limit $\Lambda > 60$ MeV for the scale parameter of $SU(2)$ YMT.
We also calculated the interglueball potentials in $SU(3)$ and $SU(4)$ YMTs.
The result shows that they are similar in size with the $SU(2)$ one, and the expected behavior of $1/N_c^2$ in the large $N_c$ expansion could not be clearly seen.
For the case of $SU(3)$, the result also suggests that the systematics due to the discretization is not negligible.
This motivates us to perform the analysis of the interglueball scattering with the improved action which has better rotational symmetry than the standard plaquette action that we used in this work.

The calculations were carried out on SX-ACE at RCNP/CMC of Osaka University.

\end{document}